# Graphene bubbles with controllable curvature

T. Georgiou[1], L. Britnell[1], P. Blake[2], R. Gorbachev[1], A. Gholinia[3], A. K. Geim[1], C. Casiraghi[4,5], K. S. Novoselov[1,*]

[1]School of Physics and Astronomy, University of Manchester, Manchester, UK
[2]Manchester Centre of Mesoscience and Nanotechnology, University of Manchester, Manchester UK
[3]School of Materials, University of Manchester, Manchester, UK
[4]Physics Department, Free University Berlin, Germany
[5]School of Chemistry and Photon Science Institute, University of Manchester, Manchester, UK

**Raised above the substrate and elastically deformed areas of graphene in the form of bubbles are found on different substrates. They come in a variety of shapes, including those which allow strong modification of the electronic properties of graphene. We show that the shape of the bubble can be controlled by an external electric field. This effect can be used to make graphene-based adaptive focus lenses.**

Graphene is a remarkable material with a number of unique properties[1,2]. Recently, most of the research was concentrated on its electronic[3] and optical properties[4-6], which is indeed justified as graphene is expected to have a big impact in the semiconductor industry. However, there are a number of other characteristics of this 2D crystal which are unique or far superior to those in other materials. For instance, graphene is impermeable to gas[7], it is very elastic (can be stretched up to 20%)[8,9] and optically transparent[4]. These properties allow creating graphene bubbles of various shapes. It has been shown that the electronic properties of such curved graphene are strongly modified by strain, which might be used for band-structure engineering[10-13]. In this letter we demonstrate a possible use of circular bubbles with controllable shape for optical lenses with variable focal length.

Many modern optical systems require adaptive focus lenses. Such lenses are highly sought after in the mobile phone industry for use in cameras, with applications also in webcameras, automatic beam steering and auto-focusing eyeglasses. Miniaturization of components makes it difficult to decrease the size of conventional, mechanically moving elements, as this increases their cost and fragility and decreases the performance, rendering them unacceptable for low-cost consumer products. The current technologies for tunable-focus lenses range from liquid-crystal[14,15] to fluid-filled lenses[16,17]. However, their fabrication is rather complex and usually includes two or more liquids or a liquid crystal layer sandwiched between two ITO layers that serve as the electrodes. Graphene-based optics can provide simplified fabrication methods, and as fabrication of large scale graphene reaches maturity, scalability should not be a problem.

Graphene flakes were produced by micro-mechanical exfoliation of single-crystal graphite[18,19] and deposited on oxidized silicon wafer, previously cleaned with oxygen plasma. The bubbles were identified by optical microscopy[20-22]. Monolayer and bilayer graphene flakes were identified by Raman

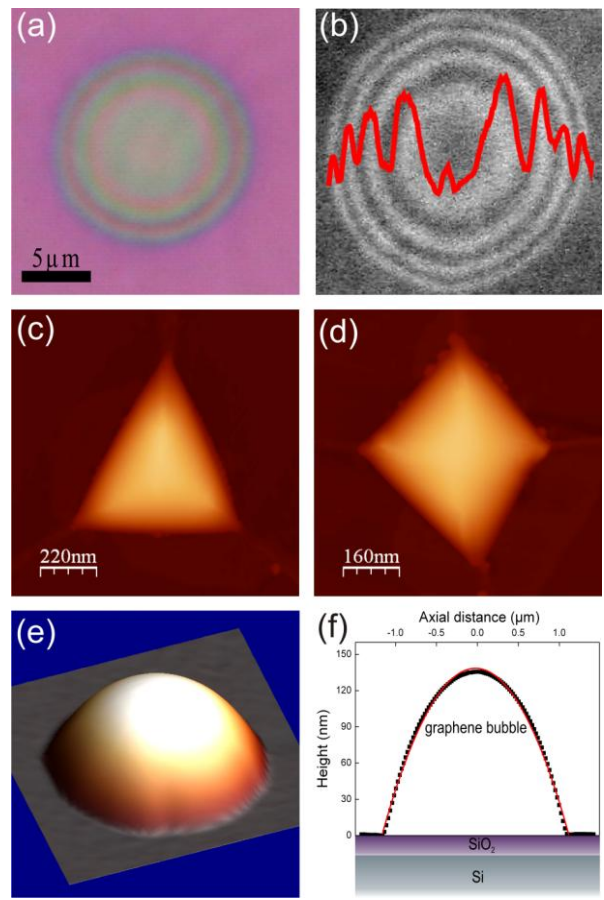

Figure 1 (a) Optical micrograph under white light of a graphene bubble deposited on silicon. The Newton rings are well visible; (b) Optical micrograph under monochromatic light of a graphene bubble deposited on Silicon: the Newton rings appear as a series of bright and dark fringes. Inset: intensity profile; (c, d) AFM topography scan of triangular and square bubbles on BN substrate; (e) 3μm x 3μm topography AFM scan of a bubble; (f) Schematic diagram of the substrate and cross-section of the bubble, obtained by AFM. Squares – experimental data, the solid line is the fit of the cross-section by assuming spherical shape.

*email: kostya@manchester.ac.uk

spectroscopy[23]. A Renishaw Raman spectrometer equipped with 514 nm laser line and power well below 1mW was used. The shape of the bubble was investigated by Atomic Force Microscopy (AFM) in tapping mode.

Bubbles are regularly found at the silicon-oxide/graphene interface in large flakes (lateral size above 0.1mm), obtained by micromechanical cleavage[24]. The origin of these bubbles is not completely understood: they probably arise from air or hydrocarbon residuals trapped between graphene and the substrate. The bubbles are well visible under the optical microscope: a monolayer reflects enough light to generate characteristic fringes of different colors, so-called Newton rings[25]. Such graphene blisters come in various shapes and sizes (from a few tens of nanometers to tens of microns): a bubble with triangular cross-section can be very interesting because it has been shown that trigonal deformation of a graphene membrane can generate a quasiconstant pseudo-magnetic field, which leads to the opening of a sizable gap in the electronic spectrum[10-13]. On the other side, circular bubbles can find use in optics. Here we show that it is possible to control the curvature of a spherical graphene bubble by applying an electric field. This allows making a graphene-based lens with adaptive focus.

Figure 1 (a) shows the optical micrograph of a single-layer bubble deposited on oxidized silicon. The Newton rings are well visible: they are the result of light interference of the reflected and transmitted light between the spherical surface of the bubble and the reflective silicon substrate. Under monochromatic light, the Newton rings appear as a series of bright and dark fringes, Figure 1(b).

We have also observed small bubbles when micromechanically cleaved graphene is deposited on very flat substrates, such as crystalline Boron Nitride (BN)[26]. Figure 1 (c-d) shows the AFM images of two bubbles on BN with square and triangular cross-sections.

Figure 1(e) shows a 3D AFM image of a bubble on a silicon substrate. The height profile of the bubble can be accurately fitted by a sphere, as predicted by simple elasticity theory[27], Figure 1 (f).

Using the fit from Figure 1(f), we can calculate the strain in the bubble: assuming no strain when graphene is in contact with the substrate, i.e. when there is no bubble, the strain is given by $\varepsilon = (l-a)/a$, where $l$ is the length of the arc and $a$ is the width of the bubble. We found that the strain in the graphene bubble is 1%. This is in agreement with Raman spectroscopy measurements on graphene bubbles, which indicate a strain of 1%[28].

We prepared several graphene devices which contained bubbles of about 5 to 10 μm in width: contacts are made by evaporating Ti/Au (5nm/50nm) through a stencil mask to avoid the use of resists and hence reducing contamination. This enables us to apply a gate voltage ($V_g$) in back-gating configuration. Optical pictures of the bubble, with a 50X objective and a narrow-band green filter with λ=510nm, were recorded while tuning $V_g$ from -35V to +35V.

Figure 2(a) shows the intensity profiles of the rings as a function of $V_g$. This Figure shows that the position of the maxima and minima of the intensity profile changes with $V_g$. In particular, at negative voltage, the radius of the rings decreases, i.e. the bubble shrinks. Figure 2(b) shows the radius of each of the Newton rings as a function of $V_g$.

Since the position of the Newton rings depends on the local height of the bubble, we can now reconstruct the shape of the bubble at every $V_g$. We calculate the height corresponding to every ring by using a four-interface interference Fresnel law-based model for incident light of *λ=510nm* wavelength[20]. We found that the bright fringes occur at the bubble height of *866, 612, 355* and *102nm*, while the dark fringes occur at *740, 482* and *229nm*. The reconstructed shape of the bubble at every *Vg* is shown in Figure 3. This Figure shows that the radius of curvature of the bubble changes from 31.3μm at 0V to 26.7μm at -35V.

One can create a lens out of such bubbles either by filling them up with some liquid or by

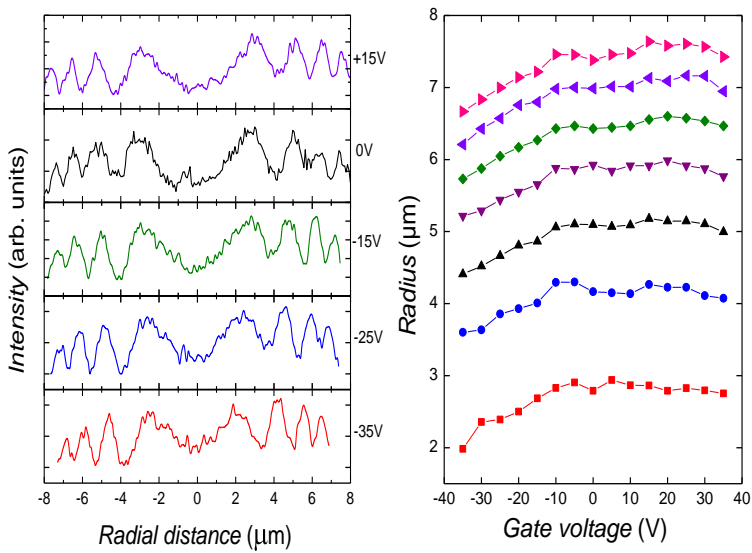

Figure 2 (a) Intensity profile of the Newton rings across the width of the bubble for different gate voltages. Maxima (minima) correspond to bright (dark) fringes, respectively. (b) Radii of bright and dark Newton rings as a function of gate voltage. The radius of the rings is clearly decreasing for negative $V_g$.

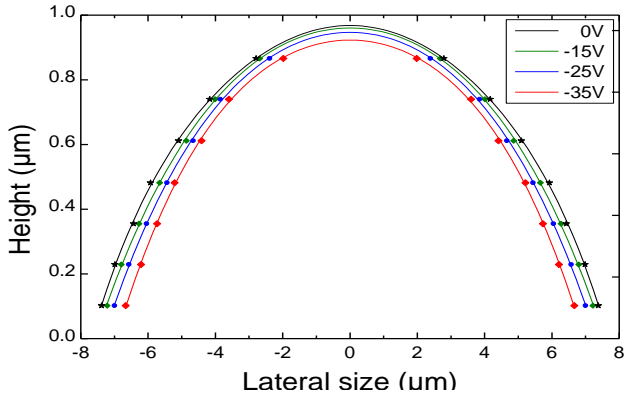

Figure 3 Reconstructed shape of the bubble for $V_g$=0V, -15V, -25V, -35V. The application of a gate voltage results in a change in the bubble's shape.

covering a bubble with a flat layer of liquid (in both cases the liquid should have high refractive index). By considering such system as a thin lens, the focal length, *f*, can be approximated as: $f \sim R/(n_2-n_1)$, where *R* is the radius of curvature of the bubble and $n_1$ and $n_2$ the refractive index of the material inside and outside the bubble, respectively. Thus, we achieved a tuning ratio of the focal length of $\Delta f/f \sim 15\%$ at $V_g = -35V$.

In first approximation, one would expect the bubbles to have the largest size at zero applied bias, while they should shrink for both positive and negative gate voltages because of electrostatic interaction: the charge on the gate would be exactly opposite to that on graphene and thus create an attractive potential. In our experiment, however, the point of the maximum size is shifted to approximately +15V and the bubble does not change shape at positive voltage, Figure 2(b). This indicates that our graphene layer is p-doped, as observed also by Raman Spectroscopy[28]. We attribute this to the presence of a water layer trapped between graphene and the substrate, which is known to p-dope graphene[29]. Since graphene is hydrophobic, such water layer is not present inside the bubble. Nevertheless, one expects the edge of the bubble to still be positively charged due to finite screening length.

In conclusion, we have shown that graphene, being very elastic and impermeable to gases, can form bubbles of various shapes and sizes. In particular, circular bubbles can be used in optics as adaptive-focus lenses. We have shown that it is possible to control the curvature of the bubble by electrostatic interaction produced by an external electric field. Graphene-based lenses have high transmittance throughout the visible range, they are light, robust and require low operating voltage.